\documentclass[twocolumn,prc,showpacs,amsmath,amssymb,superscriptaddress]{revtex4-1}
\usepackage{graphicx,color}
\usepackage{amssymb}
\usepackage{amsmath}
\usepackage{latexsym}
\usepackage{amsxtra}
\usepackage{amscd}
\usepackage{braket}
\usepackage{verbatim}
\def\vec#1{\boldsymbol{#1}}
\newcommand{\nuc}[2]{$^{#1}$#2}
\newcommand{\ko}[3]{$^{#1}$#2($-$#3)}

\newcommand{\cg}[1]{(#1)}

\newlength{\figwidth}

\begin{document}
\setlength{\figwidth}{0.98\columnwidth}

\title{Microscopic two-nucleon overlaps and knockout reactions from $^{12}$C}

\author{E.~C. Simpson}
\affiliation{Department of Physics, Faculty of Engineering and Physical
Sciences, University of Surrey, Guildford, Surrey GU2 7XH, United Kingdom}

\author{P. Navr\'atil}
\affiliation{TRIUMF, 4004 Wesbrook Mall, Vancouver, British Columbia,
V6T2A3, Canada}

\author{R. Roth}
\affiliation{Institut f\"{u}r Kernphysik, Technische Universit\"{a}t
Darmstadt, 64289 Darmstadt, Germany}

\author{J.~A. Tostevin}
\affiliation{Department of Physics, Faculty of Engineering and Physical
Sciences, University of Surrey, Guildford, Surrey GU2 7XH, United Kingdom}
\affiliation{National Superconducting Cyclotron Laboratory, Michigan State
University, East Lansing, Michigan 48824, USA}

\date{\today}

\begin{abstract}
The nuclear structure dependence of direct reactions that remove a
pair of like or unlike nucleons from a fast $^{12}$C projectile
beam are considered. Specifically, we study the differences in the
two-nucleon correlations present and the predicted removal cross
sections when using $p$-shell shell-model and multi-$\hbar\omega$ no-core
shell-model (NCSM) descriptions of the two-nucleon overlaps for the
transitions to the mass $A$=10 projectile residues. The NCSM calculations
use modern chiral two-nucleon and three-nucleon (NN+3N) interactions.
The $np$-removal cross sections to low-lying $T$=0, $^{10}$B final
states are enhanced when using the NCSM two-nucleon amplitudes. The
calculated absolute and relative partial cross sections to the low
energy $^{10}$B final states show a significant sensitivity to the
interactions used, suggesting that assessments of the overlap functions
for these transitions and confirmations of their structure could be made
using final-state-exclusive measurements of the $np$-removal cross sections
and the associated momentum distributions of the forward travelling
projectile-like residues.
\end{abstract}

\maketitle

\section{Introduction}
Properties of the wave functions of pairs of nucleons in a mass $A+2$
projectile can be probed using sudden two-nucleon removal reactions, that
exploit fast, surface-grazing collisions of the projectile with a light
target nucleus. The sensitivity is to the wave functions of the nucleons
at and near the surface of the projectile. When combined with $\gamma$-decay
spectroscopy, partial cross sections of populated final-states in the mass
$A$ reaction residues can be determined. This direct reaction mechanism,
its cross sections, and their distributions with the momenta of the forward
travelling residues, are now being exploited as a spectroscopic tool in
studies of the evolution of nucleon single-particle structure near the
Fermi surfaces of some of the most exotic nuclei; see for example Refs.
\cite{HaT03,BHS02,TBB04} and citations therein. The reaction observables
used are, currently, inclusive with respect to the final states of both
the removed nucleons and of the struck light target nucleus. More exclusive
measurements, e.g. of the light charged fragments in the final-state, may
in the future provide additional probes of the projectile structure
\cite{wimmer2}.

A detailed discussion of the two-nucleon removal reaction mechanism, its
eikonal reaction-dynamical description, and the cross sections and their
momentum distributions, in the case of high-energy two-nucleon removal
from \nuc{12}{C}, was presented in Ref. \cite{previous}. A feature of
this model description is that the removal cross sections involve only
elastic interactions of the projectile residues with the target but sums
of contributions from both elastic and inelastic interactions of one or
both nucleons with the target \cite{ToB06}. New data, for the $sd$-shell
nucleus \nuc{28}{Mg} and the \ko{28}{Mg}{2p} reaction, have quantified
these different contributions experimentally \cite{wimmer1}, and have
confirmed that the relative importance of these different processes
to the cross sections are consistent with the predictions of the eikonal
dynamical model. This has provided an important additional test of the
reaction model. In the earlier work for \nuc{12}{C} \cite{previous},
the theoretical comparisons used the sums of these removal contributions
and $p$-shell ($0\hbar\omega$ shell-model) structure calculations were
used to construct the required $\langle ^{10}\text{X}(J_f^\pi,T)|^{12}
\text{C} \rangle$ two-nucleon overlaps. The {\sc pjt} \cite{pjt} and
the {\sc wbp} \cite{wbp} shell-model effective-interaction Hamiltonians
were used.

Key elements of that analysis are also relevant here:\\
\indent
(i) The reaction is geometrically selective \cite{STB09b} and the
two-nucleon removal cross sections will be enhanced if the projectile
ground-state has components with pairs of nucleons with strong spatial
correlations (localization).\\
\indent
(ii) The available experimental cross section data, from high-energy
primary-beam measurements, are inclusive with respect to the populated
bound states of the residues following $np$, $nn$ and $pp$ removal
\cite{KLC88,LGH75}. The data, at three energies, reveal a significant
enhancement of the ratio of unlike-pair yields, $\sigma_{-np}$, to
those for the like-nucleon pairs, $\sigma_{-nn}$ and $\sigma_{-pp}$.
This enhancement is significantly greater than that expected based
simply on the numbers of available 2N-pair combinations (i.e. a factor
of 8/3). For example, the experimental $\sigma_{-np}:\sigma_{-nn}$ ratio
was 8.54 for the data set with a \nuc{12}{C} beam of energy 2.1 GeV
per nucleon. Some (but not all) of this enhancement could be explained
as due to the pair-correlations already generated in $0\hbar\omega$
$p$-shell-model overlap functions and because a larger fraction of the
$nn$-removal strength leads to unbound \nuc{10}{C} final states. However,
the experimental $\sigma_{-np}$ remained factors of 1.45 to 2.2 larger
than the theoretical model calculations for the three available data
sets \cite{KLC88,LGH75}. Table \ref{12C}, reproduced from Ref.
\cite{previous}, shows both the $p$-shell-model results and data.

\begin{table*}[t]
\caption{Calculated and experimental cross sections for two-nucleon
removals from $^{12}$C, for projectile energies of 250, 1050 and
2100 MeV per nucleon, from \cite{previous}. All cross sections are in mb.
The calculations use $p$-shell shell-model wave functions from the
{\sc wbp} effective interaction.\label{12C} }
\begin{ruledtabular}
\begin{tabular}{c|ccc|ccc|ccc}
Energy&\multicolumn{3}{c|}{\nuc{10}{Be}} &
\multicolumn{3}{c|}{\nuc{10}{C}}&
\multicolumn{3}{c}{\nuc{10}{B}} \\
MeV/u&$\sigma_{-2N}$ &$\sigma_{exp}$ & $\sigma_{exp}/\sigma_{-2N}$ &
$\sigma_{-2N}$ &$\sigma_{exp}$ & $\sigma_{exp}/\sigma_{-2N}$ &
$\sigma_{-2N}$ &$\sigma_{exp}$ & $\sigma_{exp}/\sigma_{-2N}$ \\
\hline 250\  \cite{KLC88} & 7.48 &5.88$\pm$9.70& 0.79$\pm$1.30
&5.80&5.33$\pm$0.81& 0.92$\pm$0.14
& 21.57 &47.50$\pm$2.42 & 2.20$\pm$0.11\\
1050 \cite{LGH75}& 6.62 &5.30$\pm$0.30& 0.80$\pm$0.05 &5.13
&4.44$\pm$0.24&0.87$\pm$0.05
& 19.27 &27.90$\pm$2.20 & 1.45$\pm$0.11\\
2100 \cite{LGH75} &6.52&5.81$\pm$0.29& 0.89$\pm$0.04 &5.04
&4.11$\pm$0.22&  0.82$\pm$0.04
&19.02&35.10$\pm$3.40 & 1.84$\pm$0.18\\
\end{tabular}
\end{ruledtabular}
\end{table*}
\indent
(iii) The shapes and widths of the reaction residues' momentum
distributions have both diagnostic and spectroscopic value, being
indicative of the total angular momentum, $I$, the total orbital
angular momentum $L$, and hence, with $(LS)I$ coupling, also the
total spin $S$ carried by the removed nucleon pair \cite{SiT10}.\\
\indent
(iv) The calculated cross sections for the $T$=1 states common to all three
residues, namely the first 0$^+$ and 2$^+$ states, are essentially
equal. Minor differences in the calculations arise from the small
differences in the empirical separation energies for each system.
Unlike the $np$-removal case, these calculated inclusive $T$=1
cross sections were reasonably consistent with and were fractionally
larger than the data values for $\sigma_{-nn}$ and $\sigma_{-pp}$,
suggesting that the deficit in the theoretical cross sections in
the $np$ channel reflects, primarily, a failure of our description
of the overlaps $\langle ^{10}\text{B}(J_f^\pi,T=0)|^{12}\text{C}
\rangle$ for the transitions to the $T$=0, \nuc{10}{B} final states.

\begin{figure}[b]
\includegraphics[width=0.8\figwidth]{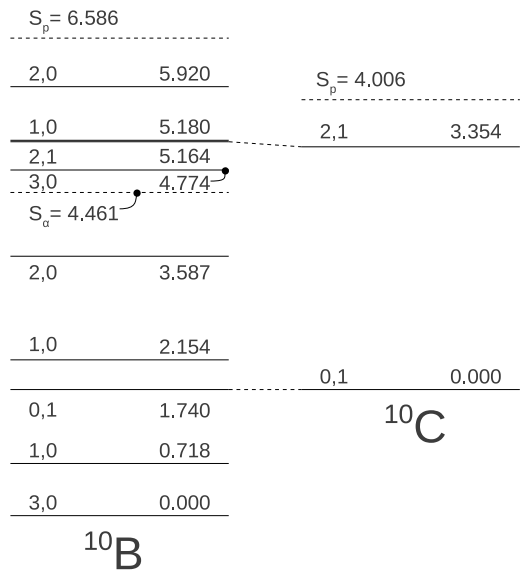}
\caption{States of the mass $A=10$ residues populated in the
two-nucleon removal reactions considered. The spin and isospin
labels ($J_f$,$T$) are indicated. All states included are of
positive parity. Levels assumed to be part of the isospin multiplet
are connected by dashed lines. The lowest particle thresholds are
also indicated. States above the $\alpha$-particle threshold in
\nuc{10}{B} are expected to decay via $\alpha$ emission, with the
exception of the 5.164 MeV, $T$=1, $J_f^\pi=2^+$ state which has
an 84\% $\gamma$-decay branch. \label{fig:levels}}
\end{figure}

Here we exploit the eikonal reaction model in the isospin formalism
\cite{TPB04,ToB06,STB09b} for the removal of the like ($T$=1) and
unlike ($T$=0,1) pairs of nucleons from \nuc{12}{C}. In particular,
we will investigate the effect on reaction observables when using
{\em ab-initio} multi-$\hbar\omega$ no-core shell-model (NCSM)
descriptions for the two-nucleon overlaps. We contrast these with
the earlier $0\hbar\omega$ $p$-shell shell-model results of
Ref. \cite{previous}. We discuss the $np$ and $nn$ removal channels
to \nuc{10}{B} and \nuc{10}{C}. These channels share the
same $T=1$, $0_1^+$ and $2_1^+$ final states, see e.g. Fig. \ref{fig:levels},
whereas the $pp$ channel contains two additional $T=1$ states.

In Section \ref{sec:carbon} we reiterate some specific features of the
reaction description for $^{12}$C projectiles. The necessary formalism
has been presented elsewhere \cite{ToB06,STB09b} to which readers are
referred. We follow the notations used in these earlier works. The
chiral effective field theory (EFT) two- and three-nucleon (NN+3N)
interactions and the microscopic no-core shell-model (NCSM) calculations
used to construct the improved overlap functions will be discussed
in section \ref{sec:struct}. These include calculations in which the
chiral 3N interaction is switched off in an attempt to understand
the impact on observables of these 3N interaction terms in the
starting Hamiltonian. The new results and predictions for \nuc{12}{C}
reaction observables are analyzed in section \ref{sec:results} and
a summary is presented in section \ref{sec:summary}.

\section{Carbon-induced reactions \label{sec:carbon}}
Consideration of two-nucleon removal from $^{12}$C is valuable given
the availability of both conventional shell-model and ab-initio NCSM
structure descriptions. The residual nuclei $^{10}$C, $^{10}$B (and
$^{10}$Be) were also extensively studied and so establish a valuable
benchmark. In addition, the existing experimental cross section
measurements \cite{LGH75,KLC88}, although inclusive with respect
to the residue final states, have relatively small quoted
uncertainty and were taken at high energies where the eikonal model
description of the reaction dynamics is most reliable. These data
were obtained for reactions of a carbon beam and carbon target at 250,
1050 and 2100 MeV per nucleon incident energies. As mentioned above,
these data (and related data for other light projectile nuclei) show
a significant enhancement in their $T$=0,1, $np$-removal production
cross sections over those with $T$=1, the $nn$ and $pp$ removal cases
(see e.g. Table \ref{12C}). This observed enhancement is of particular
interest as a potential signal and a measure of strong $np$-correlations
at the nuclear surface.

The primary motivation for the present study is the implementation and
first assessment of improved microscopic descriptions of the two-nucleon
overlap functions and their implications for $np$-correlations and
the calculated reaction yields and observables. The relevant final
states of the $A$=10 residual nuclei and their spins and isospins are
shown in Fig. \ref{fig:levels}. The known spectrum of low-lying states
in \nuc{10}{B} also contains several negative parity states. These
are not expected to be populated in the two-nucleon removal reaction
mechanism. More exclusive measurements would be needed to confirm
this expectation.

The isospin formalism developed in Refs. \cite{ToB06,STB09b} is
used here. The description of the reaction and the parameters used
for the \nuc{12}{C} and mass $A$=10 residue densities, etc. are the
same as were discussed and tabulated in Ref. \cite{previous}. The
key approximations were as follows: (a) The removal of nucleons is
sudden, their co-ordinates frozen during the short timescale of the
relativistic, surface-grazing reactions. (b) The no-recoil (heavy
residue) approximation is made \cite{ToB06}. The inclusion of core
recoil affects only the diffractive-stripping terms of the reaction
mechanism whose contributions are a significantly smaller fraction
of the cross section ($\approx$30\%) than the two-nucleon stripping
terms, for which recoil is not an issue. Thus, recoil can play, at
most, a minor role on computed cross sections. It is, in any case,
the fractional changes of the cross sections between the NCSM and
the $p$-shell calculations, and between $T$=0 and $T$=1 configurations
that are of most interest and significance to this work.

The primary difference here is a now considerably-extended set
of two-nucleon amplitudes (TNA) $C_{\alpha}^{IT}$ arising from the
multi-shell set of available 2N-configurations $\alpha \equiv [\beta_1,
\,\beta_2$]. Here the index $\beta=(n \ell j)$ denotes the spherical
quantum numbers of each active single-particle state in the model space.
We evaluate the cross sections for transitions from the projectile
initial (ground) state $i$, with spin $(J_i, M_i)$, to particular
residue final states $f$, with $(J_f, M_f)$. The all-important
two-nucleon overlap function for removal of two nucleons 1 and 2 is
\begin{align}
\Psi_i^{(F)} = &\sum_{I\mu{T}\alpha}C_{\alpha}^{IT} \cg{I{\mu}J_fM_f|
J_iM_i} \nonumber\\ &\cg{T{\tau}T_f\tau_f|T_i\tau_i} \; [\,\overline{
\psi_{\beta_1}(1)\otimes \psi_{\beta_2}(2)}\,]_{I \mu}^{T\tau}\ .
\label{eqn:overlap}
\end{align}
Here $J_i^\pi$=$0^+$ and thus $J_f$=$I$, the 2N total angular momentum.
We discuss later the use of Woods-Saxon or harmonic oscillator radial
functions for the single-particle orbitals $\psi_{\beta}$. The 2N
correlations under discussion arise in Eq. (\ref{eqn:overlap}) from:
(a) somewhat trivially, the antisymmetry and angular momentum coupling
of nucleon pairs, and (b) the possible coherent pair-enhancement arising
from the weights and phases of the TNA that contribute to each $J_f$
final state. The latter and their sensitivity to: (i) the NN+3N effective
interactions used, and (ii) the NCSM model-space dimensions, are
studied here.

Importantly, unlike the earlier $p$-shell-only analysis, the NCSM wave
functions and overlaps now include active single-particle orbitals of
both parities, including up to $N_{max}=6$ major oscillator shells. The
angular correlation function of the two nucleons thus contains both
even and odd Legendre polynomial terms in the angular separations of
the two nucleons \cite{Pin84,SiT10} with the likelihood of a greater
spatial localization of pairs at the nuclear surface. The degree of
such increased correlations is discussed further in the results section
where we also show the two-nucleon joint-position probabilities in the
impact parameter plane, $\mathcal{P}_{J_f} (\vec{s}_1, \vec{s}_2)$, as
were defined in Ref.\ \cite{SiT10}. These display the two-nucleon
spatial correlations as delivered by the projectiles and as are seen
by the target nucleus that induces the reaction.

Nucleon removal may occur via either elastic (diffraction) or inelastic
(stripping) interactions of nucleons with the target nucleus, the former
leaving the target in its ground state. The latter lead to cross section
contributions that are inclusive with respect to all other target final
states. Two-nucleon removal events can involve: (a) both nucleons making
inelastic collisions with the target, (b) there being one inelastic and
one elastic collision, or (c) both nucleons suffering elastic collisions.
Events (b), referred to as diffraction-stripping, are identified in the
reaction's absorption cross section but require a projection-off bound
states for the elastically interacting nucleon (for details see Refs.
\cite{ToB06,STB09b}). When truncated basis, single major shell, shell-model
calculations have been used to generate the TNA, all active single-particle
orbits in the shell were included in this projection operator. Here, when
using the NCSM calculations in a very large basis, we have limited the
projection to the $0p$-shell orbitals, as being the appropriate bound
states set. Purely elastic 2N-removal events (c) were estimated, as
previously \cite{ToB06}, from the stripping and diffractive-stripping
cross sections. Typically, they contribute to the cross sections at
the level of $<5\%$.

The absorptive eikonal S-matrices, that largely determine the volume
of the two-nucleon overlap function that is sampled in the reaction, were
calculated by folding the target, nucleon, and reaction residue point-nucleon 
densities with a zero-range effective nucleon-nucleon interaction. Further
details may be found in Ref. \cite{previous}. Our primary interest
is in the $np$ removal cross sections. Since both the 0$_1^+$ and
2$_1^+$ $T=1$ states are populated in both \nuc{10}{B} and \nuc{10}{C}
we may also extract the cross section for the $nn$ channel. Very minor
differences in these partial cross sections will arise from (a) the
use of a \nuc{10}{B} rather than a \nuc{10}{C} S-matrix, and (b) from
the small binding energy differences. The latter are not accounted for
in the present calculations where harmonic oscillator radial wave
functions are used.

\section{NN+3N interactions and overlaps \label{sec:struct}}
The $p$-shell (0$\hbar\omega$) shell-model calculations and overlaps
were described in Ref. \cite{previous} and were computed using the
code {\sc oxbash} \cite{BEG04}. For the present work, a series of
no-core-shell-model calculations, each for a given number of
major oscillator shells, $N_{max}=$0, 2, 4 or 6, were carried out
using two chiral EFT NN+3N interaction choices, denoted {\sc ncsm1}
and {\sc ncsm2} in the following.

The calculations used interactions derived within the chiral effective field theory (EFT) approach.
In particular, the chiral N$^3$LO NN interaction of Ref.~\cite{Entem:2003ft,Machleidt:2011zz} was used with or without
the chiral N$^2$LO 3N interaction~\cite{Epelbaum:2002vt} in the local form of Ref. ~\cite{Navratil:2007zn}.
These interactions were softened by the similarity renormalization group
(SRG) technique~\cite{Glazek:1993rc,Wegner:1994_JPV,Jurgenson:2009qs}, where a unitary transformation is used to
suppress the off-diagonal matrix elements (controlled by a parameter $\Lambda$). The SRG interaction induces higher-body
interaction terms. These induced terms were kept up to the three-body level. It has been shown~\cite{Jurgenson:2010wy,Roth:2011ar}
that four- and higher-body terms are negligible for light nuclei although some evidence for four-body induced terms was observed
in $^{12}$C calculations with one of the interactions used here ({\sc ncsm2})~\cite{Roth:2011ar}.
In {\sc ncsm1} the NN+3N Hamiltonian used a 3N cutoff of 400 MeV and used parameters fitted to the $^3$H
lifetime and the $^4$He binding energy~\cite{Roth:2011vt}. In {\sc ncsm2} the 3N cutoff
was 500 MeV and the parameters were fitted to the lifetime and binding
energy of $^3$H~\cite{Gazit:2008ma}. In both cases the SRG was carried out using $\Lambda
$=1.7 fm$^{-1}$, although the {\sc ncsm2} calculations were also performed with $\Lambda$=1.88 fm$^{-1}$ to verify the SRG-$\Lambda$ independence, i.e., to confirm the unitarity of the SRG transformation. The subsequent NCSM calculations used an harmonic
oscillator (HO) basis with an angular frequency $\hbar\omega$=16 MeV.
The mass-dependent parameterizations of the oscillator frequency $\hbar\omega=45A^{-1/3}-25A^{-2/3}$,
agreeing with charge radius observations, suggest a value of
$\hbar\omega$=14.9 MeV \cite{BlM68}, in reasonable agreement with the
value used here.

In the case of the {\sc ncsm2} parameterization, the calculations were also
repeated, and denoted as {\sc ncsm3}, when the chiral 3N interaction in
the starting Hamiltonian was switched off, but with the SRG-induced 3N
effects (with $\Lambda$=1.7 fm$^{-1}$) included. Again, the HO frequency of $\hbar\omega$=16 MeV was employed.
Using these {\sc ncsm3} TNA we can make a first assessment of the impact on calculations/observables
of the inclusion, or not, of the chiral 3N interaction in the starting Hamiltonian.

\begin{figure}[htb]
\includegraphics[angle=-90,width=\figwidth,clip=true]{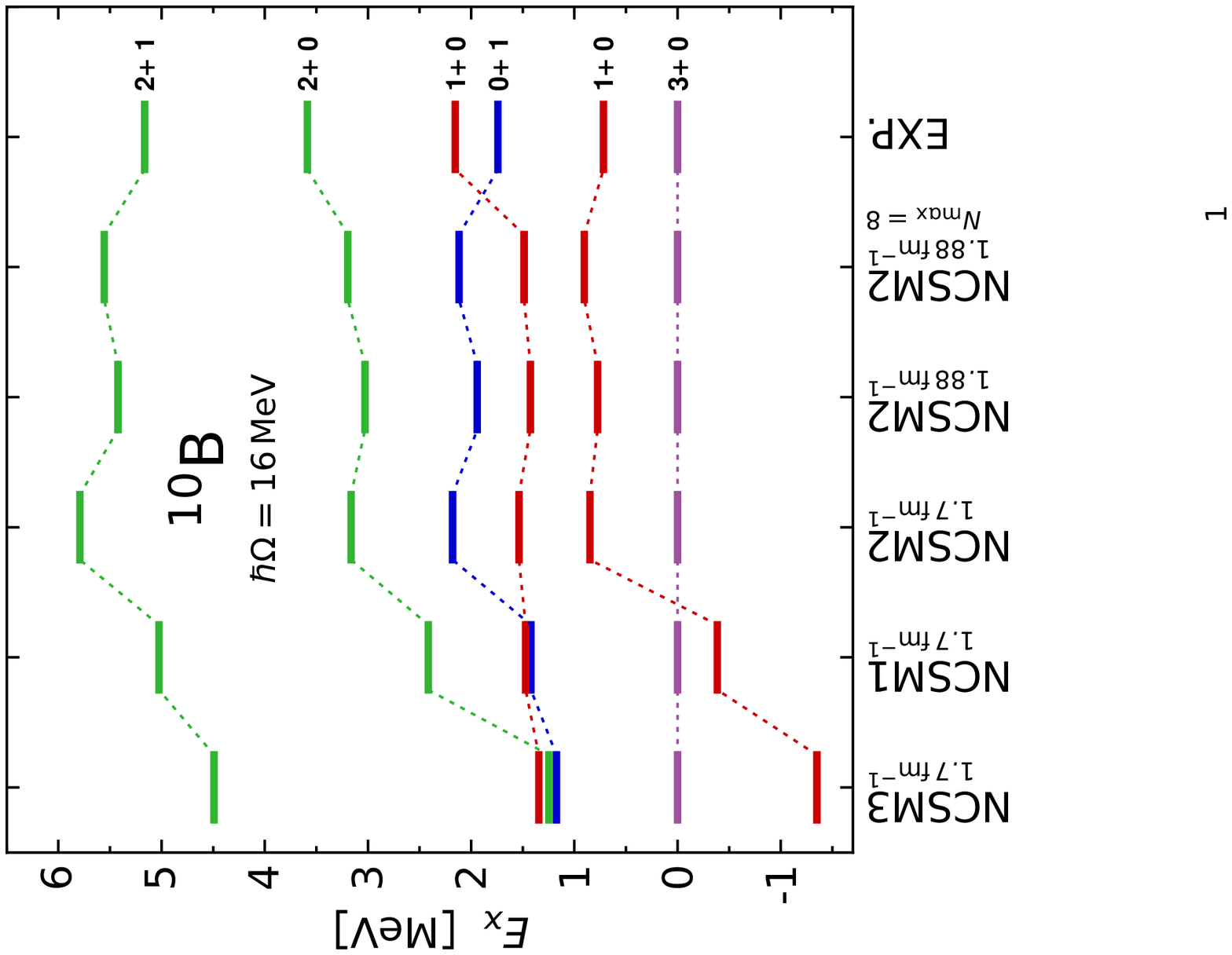}
\caption{(Color online) Experimental excitation energies of $^{10}$B are compared to the different calculations used in the present work: chiral N$^3$LO NN ({\sc ncsm3}), chiral N$^3$LO NN plus N$^2$LO 3N with the cutoff of 400 MeV ({\sc ncsm1}), and chiral N$^3$LO NN plus N$^2$LO 3N with the cutoff of 500 MeV ({\sc ncsm2}). The $N_{\rm max}{=}6$ space was used in calculations shown in the first four columns. The SRG $\Lambda$ parameter is indicated. The HO frequency of $\hbar\Omega{=}16$ MeV was used in all calculations.} \label{fig:10B_spectra}
\end{figure}

It should be noted that the $^{10}$B structure poses a particular challenge to {\it ab initio} calculations. In particular,
it had been observed that standard accurate NN potentials predict incorrectly the ground-state of $^{10}$B to be $1^+ 0$, instead of the experimental $3^+ 0$. The present calculations with the chiral N$^3$LO NN potential ({\sc ncsm3}) suffer from the same problem. Only after including the chiral N$^2$LO NNN term, with the 3N cutoff of 500 MeV {\sc ncsm2}, does one get the correct ground state spin. Interestingly, the weaker chiral N$^2$LO NNN with the 3N cutoff 400 MeV, {\sc ncsm1}, fails to invert the $1^+ 0$ and $3^+ 0$ states, also predicting the wrong $^{10}$B ground state spin. See Fig.~\ref{fig:10B_spectra} for a comparison of $^{10}$B excitation energies from different calculations used in this paper.
Also in the figure, the stability of the spectra with respect to the SRG $\Lambda$ variation and the size of the model space $N_{\rm max}$ is demonstrated for the {\sc ncsm2} case. 
The situation is somewhat reversed in $^{12}$C, where the Hamiltonian {\sc ncsm2} with the stronger 3N interaction over-binds $^{12}$C by several MeV and overcorrects the splitting of the $1^+ 0$ and $4^+ 0$ states~\cite{Roth:2011ar}. Using the weaker 3N interaction ({\sc ncsm1}) both the binding energy and excitation energy description improves. Furthermore, this Hamiltonian ({\sc ncsm1}) also describes the binding energies of oxygen and calcium isotopes~\cite{Roth:2011vt} very well. The stronger 3N interaction {\sc ncsm2}, on the other hand, provides a very good description of lighter nuclei ($A\leq 10$), resolving even long-standing analysing power problems in $p{-}^4$He scattering~\cite{Viviani:2010mf}. These observations suggest that our knowledge of the 3N interaction in particular is incomplete and additional terms, such as those at the N$^3$LO of the chiral perturbation theory, must be included. Further, the mass region of $A{=}10-12$ is ideal to test the details of the nuclear Hamiltonians.

\section{Results }\label{sec:results}
Reaction calculations were carried out using the extended TNA sets
derived from the two NN+3N effective interactions outlined in section
\ref{sec:struct}, for each of $N_{max}=$0, 2, 4 and 6, and for the
$nn$ and $np$-removal channels. In Table \ref{12C2} we first show
the calculated inclusive cross sections for the $p$-shell {\sc wbp}
and (the most complete) NCSM $N_{max}=$6 calculations at the three
energies of the available data. The NCSM calculations use harmonic
oscillator radial form factors with $\hbar \omega$=16 MeV, as used
in the NCSM basis. The experimental inclusive cross sections were
shown in Table \ref{12C}. We note that when using the {\sc pjt}
$p$-shell effective interaction the inclusive $^{10}$B yield at
2.1 GeV per nucleon was 18.73 mb \cite{previous}, as compared to
the 19.02 mb shown in Table \ref{12C2} for the {\sc wbp} case. Thus
the two $p$-shell calculations are entirely consistent in this
inclusive cross section observable. We note that the $p$-shell-model
calculations, taken from \cite{previous}, use Woods-Saxon radial wave
functions when constructing the two-nucleon overlap functions. If
$\hbar \omega$=16 MeV oscillator functions are used the calculated
cross section is reduced by $\sim7\%$ for the lowest energy $3^+$,
and by $<3\%$ for higher energy states. The smaller changes for the higher residue
excitations are due to the absence of any binding energy effect
when a single oscillator frequency is used.
\begin{table*}[tbp]
\caption{Calculated inclusive cross sections for two nucleon
removals from $^{12}$C, for projectile energies of 250, 1050 and
2100 MeV per nucleon. All cross sections are in mb. The TNAs used
were calculated using the $p$-shell and {\sc wbp} interaction, and
the {\sc nscm1} and {\sc nscm2} NN+3N interactions with $N_{max}$=6.
The experimental cross sections were tabulated in Table \ref{12C}.
\label{12C2}}
\begin{ruledtabular}
\begin{tabular}{c|ccc|ccc}
Energy
&\multicolumn{3}{c|}{\nuc{10}{C}}
&\multicolumn{3}{c}{\nuc{10}{B}} \\
MeV/u&
$\sigma_{-2N}^\text{WBP}$&$\sigma_{-2N}^\text{NCSM1}$&$\sigma_{-2N}^\text{NCSM2}$&
$\sigma_{-2N}^\text{WBP}$&$\sigma_{-2N}^\text{NCSM1}$&$\sigma_{-2N}^\text{NCSM2}$\\
\hline
250   & 5.80 & 7.10 & 8.48 & 21.57 & 28.19 & 29.91 \\
1050  & 5.13 & 6.31 & 7.44 & 19.27 & 25.26 & 26.45 \\
2100  & 5.04 & 6.22 & 7.28 & 19.02 & 25.00 & 26.01 \\
\end{tabular}
\end{ruledtabular}
\end{table*}

The cross sections, now exclusive with respect to the $A$=10 final states,
are shown in Table \ref{tbl:c12_2n} for the highest energy of 2.1 GeV
per nucleon. The two NN+3N NCSM cases (with $N_{max}=$6) and the {\sc wbp}
case are shown together with the {\sc ncsm3} case (i.e. without the chiral
3N interaction in the starting Hamiltonian). In $np$-removal to \nuc{10}{B}
the cross sections are shown
for the six positive parity gamma-decaying final states below the first
nucleon threshold. However, the first 2$^+$, $T$=1 state is known to decay
by $\alpha$-emission with an $I_\alpha=16\%$ branch. This branching has
been accounted for in the inclusive $\sigma_{-2N}$ values presented. The
cross sections for population of the three higher lying $T$=0 states (see
Fig. \ref{fig:levels}) are not included since these states are reported to
decay by $\alpha$-emission (with $I_\alpha=100\%$). It was assumed that
these states do not contribute to the $^{10}$B yields.

Table \ref{tbl:c12_2n} reveals significant sensitivity of the yields of
the low-lying $^{10}$B states to the interactions assumed. The ratio of
the $T$=0, $1_1^+$ to $3^+$ ground-state yields is reversed between
{\sc ncsm1} and {\sc ncsm2}.  The absence of the 3N interaction in the
{\sc ncsm2} starting Hamiltonian, the {\sc ncsm3} case, leads to a quite
significantly enhanced ratio of the $T$=0 $1_1^+$ to $3^+$ ground-state
cross sections when compared to the full {\sc ncsm2} Hamiltonian. To
confront these detailed model predictions requires more exclusive
measurements with good statistics.

\begin{figure}[htb]
\includegraphics[width=\figwidth]{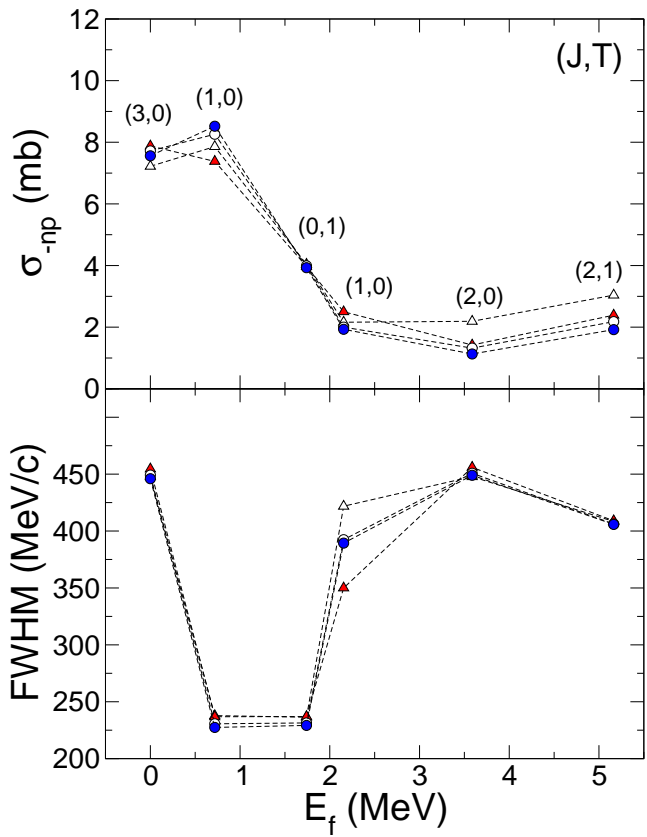}
\caption{(Color online) Calculated exclusive cross sections (upper
panel) and full width at half maximum (FWHM) widths of the momentum
distributions (lower panel) for the ground and $\gamma$-decaying final
states of the \nuc{10}{B} residues, following $np$-removal at 2100 MeV
per nucleon. The TNAs used were obtained using the {\sc ncsm1} NN+3N
starting Hamiltonian (dashed lines). Calculations are for $N_{max}$=0
(open triangles), 2 (red triangles), 4 (open circles) and 6 (blue
circles).} \label{fig:10c}
\end{figure}

\begin{figure}[htb]
\includegraphics[width=\figwidth]{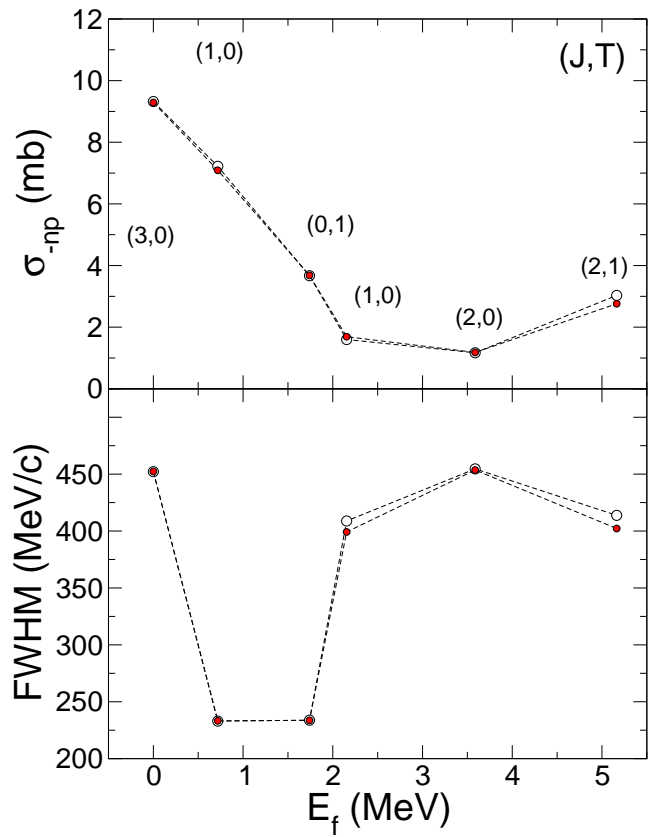}
\caption{(Color online) Calculated exclusive cross sections (upper
panel) and FWHM widths of the momentum
distributions (lower panel) for the ground and $\gamma$-decaying final
states of the \nuc{10}{B} residues, following $np$-removal at 2100 MeV
per nucleon. The TNAs used were obtained using the {\sc ncsm2} NN+3N
starting Hamiltonian (dashed lines). Calculations are for $N_{max}$=6 with the SRG $\Lambda$=1.7 fm$^{-1}$ (open circles) and $\Lambda$=1.88 fm$^{-1}$ (red circles).} \label{fig:Lambda-dep}
\end{figure}

\begin{figure}[htbb]
\includegraphics[width=\figwidth]{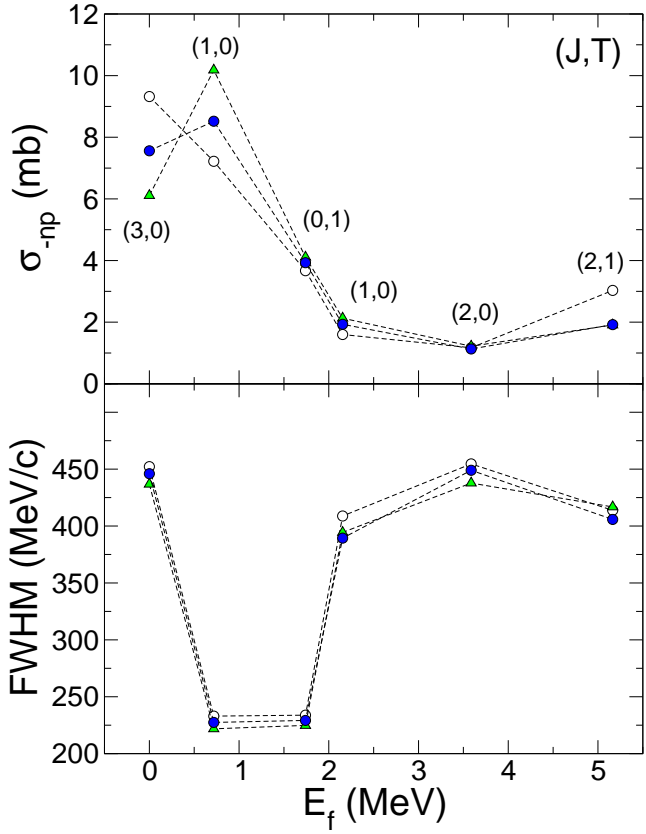}
\caption{(Color online) Calculated exclusive cross sections (upper
panel) and FWHM widths of the momentum
distributions (lower panel) for the ground and $\gamma$-decaying final
states of the \nuc{10}{B} residues, following $np$-removal at 2100 MeV
per nucleon. The TNAs used were obtained using the {\sc ncsm1} (blue
circles) and {\sc ncsm2} (open circles) NN+3N starting Hamiltonians
and the {\sc ncsm3} (triangles) NN starting Hamiltonian, all for
$N_{max}$=6.} \label{fig:10c2}
\end{figure}

\begin{figure}[htbb]
\includegraphics[width=\figwidth]{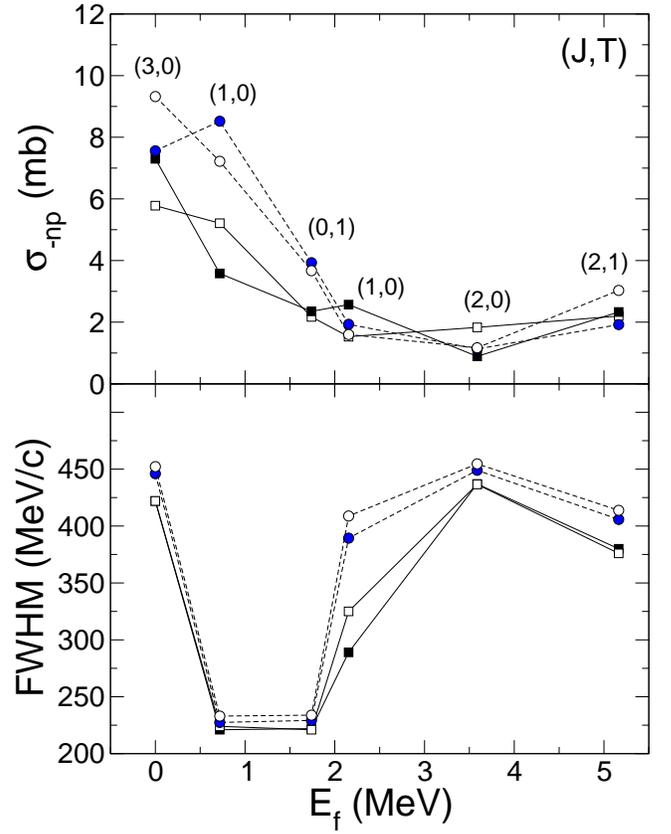}
\caption{(Color online) Calculated exclusive cross sections (upper
panel) and FWHM widths of the momentum
distributions (lower panel) for the ground and $\gamma$-decaying final
states of the \nuc{10}{B} residues, following $np$-removal at 2100 MeV
per nucleon. The TNAs used were obtained using the {\sc wbp} interaction
(solid lines and solid squares), the {\sc pjt} interaction (solid lines
and open squares) shell-model interactions and the {\sc ncsm1} (dashed
lines blue circles) and {\sc ncsm2} (dashed lines and open circles)
NN+3N Hamiltonians for $N_{max}$=6.}
\label{fig:10c3}
\end{figure}

The six contributing \nuc{10}{B} partial cross sections are also plotted
in Figs. \ref{fig:10c}, \ref{fig:10c2} and \ref{fig:10c3}. Fig. \ref{fig:10c}
shows the calculations, using {\sc ncsm1}, for $N_{max}=$0, 2, 4 and 6, and
the stabilization and the essential convergence of the calculated
partial cross sections (upper panel) and the full-width at half-maximum
widths of their momentum distributions (lower panel) with increasing
$N_{max}$. Based on this observed convergence, seen for all of the NCSM
cases, we have presented, in the main, only the final $N_{max}$=6 results.
Fig.~\ref{fig:Lambda-dep} shows $N_{\rm max}=6$ {\sc ncsm2} calculations for two different SRG 
$\Lambda$ values, $\Lambda$=1.7 fm$^{-1}$ and $\Lambda$=1.88 fm$^{-1}$. The essential independence of the cross sections and the FWHM on the SRG-$\Lambda$ confirms the unitarity of the SRG transformation, i.e., we are really investigating predictions of the initial chiral interactions. Based on the results of Ref.~\cite{Roth:2011ar}, the SRG $\Lambda$ dependence is expected to be even weaker for the {\sc ncsm1} and {\sc ncsm3} cases.
Fig. \ref{fig:10c2} shows the calculations from the {\sc ncsm1}, {\sc ncsm2}
and {\sc ncsm3} Hamiltonians, all with $N_{max}=$6. The significant variations
predicted in the widths of the momentum distributions for the different
final states and the sensitivity of the $T$=0, \nuc{10}{B} final state yields
to these effective interactions, are evident. The momentum distribution widths
from the NCSM interactions are broadly consistent, and also with those of the
conventional truncated-basis $p$-shell model interactions (shown in Fig.
\ref{fig:10c3}). They are systematically wider than the latter by $\approx$
25 MeV/c that we attribute to the incorrect asymptotic of the oscillator wave
functions. These similarities reflect the relative insensitivity of the
different $LS$-fractions in the overlap to the interactions used.
The notable exception is the second $T=0$, 1$^+$ excited state, when the width
using the NCSM interactions is somewhat wider than that from the conventional
shell-model interactions. The comparison of the NN+3N and $p$-shell shell-model
results are shown in Fig. \ref{fig:10c3}.

\begin{table}[htb]
\caption{Like and unlike two-nucleon removal cross sections (in mb)
for a \nuc{12}{C} projectile incident on a carbon target at 2100 MeV
per nucleon. The excitation energies, $E_f$, of each final state are
shown in Fig. \protect\ref{fig:levels}. The TNAs used were calculated
using (a) the $0\hbar\omega$ $p$-shell-model and {\sc wbp} interaction,
(b) the two NN+3N interactions {\sc ncsm1} and {\sc ncsm2} (see text),
and (c) the {\sc ncsm3} Hamiltonian in which the chiral 3N interaction
is turned off. Calculations (b) and (c) use the NCSM with $N_{max}=6$.
The sums show the accumulated cross sections that lead to the ground
state and the $\gamma$-decaying bound excited states of the mass $A=10$
projectile residues.
\label{tbl:c12_2n}}
\begin{ruledtabular}
\begin{tabular}{ccccccc}
Residue & $J^\pi_f$ & $T$ & $\sigma_{-2N}^\text {WBP}$ &
$\sigma_{-2N}^\text {NCSM1}$ &$\sigma_{-2N}^\text{NCSM2}$ &
$\sigma_{-2N}^\text{NCSM3}$\\ \hline
\nuc{10}{C} & 0$^+$ & 1 & 2.30 & 3.93 & 3.67 & 4.11   \\
            & 2$^+$ & 1 & 2.74 & 2.29 & 3.61 & 2.27  \\
\cline{2-7}
Inclusive\ \ \ \ &  &&5.04  &6.22 & 7.28 & 6.38  \\
Experiment& &&      &4.11$\pm$0.22  &   \\
\hline
\nuc{10}{B}
& 3$^+$    & 0& 7.30 & 7.56 & 9.32 &  6.11\\
& 1$^+$    & 0& 3.58 & 8.52 & 7.22 & 10.18\\
& 0$^+$    & 1& 2.35 & 3.93 & 3.67 &  4.11\\
& 1$^+$    & 0& 2.57 & 1.93 & 1.60 &  2.13\\
& 2$^+$    & 0& 0.89 & 1.13 & 1.17 &  1.22\\
& 2$^{+a}$ & 1& 2.33 & 1.92 & 3.03 &  1.91\\
\cline{2-7}
Inclusive\ \ \ \ & & &19.02 & 25.00 & 26.01 & 25.66\\
Experiment&& &      &35.10$\pm$3.40 &     \\
\end{tabular}
\end{ruledtabular}
\footnotetext[1]{state decays by $\alpha$ emission with a 16\% $\alpha$-branch.}
\end{table}

The final-state inclusive cross sections for the {\sc ncsm} interactions are
all larger than those from the truncated-space {\sc wbp} interaction.
All three {\sc ncsm} interactions result in similar summed cross sections to the
lowest two $T=0$ states, between 16 and 17 mb, but the distribution of this
strength between the two states shows significant variations. The cross
sections to the $T=1$, 0$^+$ state is also larger when using the {NCSM}
interactions with variations between the predictions of the interactions,
thus reinforcing the need for final-state-exclusive cross section
measurements. The detailed nature of the sensitivity of the TNA, and hence
the final-state branching ratios and cross sections, to the details of the
interactions is complex, but precise measurements could provide a path
to probe this sensitivity and toward constraining the underlying
interactions.

The ratios of the theoretical model and experimental cross sections,
$R_s (2N)=\sigma_{exp}/\sigma_{th}$ are $R_s(2n)=0.66$ and $R_s(np)=1.40$
for the {\sc ncsm1} interaction and $R_s(2n)=0.56$ and $R_s(np)=1.35$ for
{\sc ncsm2}. Calculations for two-nucleon removal in exotic $sd$-shell
isotopes typically overestimate the experimental observations by a factor
of two with $R_s(2N)=0.5$ \cite{ToB06}. As mentioned above, we do not
include core recoil and center-of-mass effects for the structure amplitudes,
so do not draw specific conclusions about these absolute values of $R_s(2N)$.
However, both effects should be independent of the reaction channel, and
the former will not affect the strongest mechanism, i.e. 2N-stripping.

To consider the impact of the significantly larger model spaces introduced
using the NCSM amplitudes, it is meaningful to consider the cross section
ratios $\sigma_{-2N}^{WBP}/\sigma_{-2N}^{NCSM}$, i.e., the relative enhancement
of a particular channel when moving from the truncated basis {\sc wbp} to
the NCSM interactions.  For the {\sc ncsm1} interaction, this ratio is 0.81
for the $nn$ channel and 0.76 for the $np$ channel.  The corresponding numbers
for the {\sc ncsm2} interaction are 0.69 and 0.73. Evidently, the use of the
larger basis amplitudes set enhances the cross section.

A significant fraction of these differences results from changes in the
$p$-shell amplitudes, as are shown in Table \ref{amplitudes}, the simplest
case being the $3^+$ ground state, where only a single $p$-shell configuration
contributes. The $[0p_{3/2}]^2$ TNA vary, depending on the different
interactions, and the magnitude of these $p$-shell amplitudes is a key
factor in the cross section changes observed.

For the two $T=0$, $1^+$ states, a mix of $p$-shell configurations now
contribute, with the overall magnitude of the TNA and their relative
strengths and phases changing. The relative strengths of the three
configurations are broadly consistent across the {\sc ncsm} interactions,
but are different from the truncated-basis {\sc wbp} interaction.
In particular the $[p_{3/2}]^2$ TNA are different, with some apparent
shift of strength from the first to the second $1^+$  state, when compared
to the {\sc ncsm} interactions.
\begin{table}[t]
\caption{p-shell two-nucleon amplitudes for the {\sc wbp}, {\sc ncsm1}, {\sc ncsm2} and
{\sc ncsm3} interactions.  \label{amplitudes} }
\begin{ruledtabular}
\begin{tabular}{cc|ccc}
Interaction & $(J^\pi,T)$ & $[p_{1/2}]^2$ & $[p_{1/2}][p_{3/2}]$ & $[p_{3/2}]^2$ \\
\hline
{\sc wbp}   & $(3_1^+,0)$ & -             & -                    & $~1.976$ \\
            & $(1_1^+,0)$ & $-0.011$      & $~0.979$             & $~0.699$  \\
            & $(1_2^+,0)$ & $~0.363$      & $~0.229$             & $-1.134$  \\
\hline
{\sc ncsm1} & $(3_1^+,0)$ & -             & -                    & $~1.913$ \\
            & $(1_1^+,0)$ & $-0.220$      & $~1.034$             & $~1.197$  \\
            & $(1_2^+,0)$ & $~0.611$      & $~0.376$             & $-0.835$  \\
\hline
{\sc ncsm2} & $(3_1^+,0)$ & -             & -                    & $~2.213$ \\
            & $(1_1^+,0)$ & $-0.255$      & $~0.863$             & $~1.307$ \\
            & $(1_2^+,0)$ & $~0.470$      & $~0.500$             & $-0.814$ \\
\hline
{\sc ncsm3} & $(3_1^+,0)$ & -             & -                    & $~1.644$\\
            & $(1_1^+,0)$ & $-0.224$      & $~1.137$             & $~1.205$ \\
            & $(1_2^+,0)$ & $~0.740$      & $~0.332$             & $-0.719$ \\
\end{tabular}
\end{ruledtabular}
\end{table}
In these cases there is interference between the different configurations that
make it less transparent what one expects from the different interactions. Changes
in the $p$-shell configurations will account for a part of the changes in cross
sections shown in Table \ref{tbl:c12_2n} and the predicted momentum distribution
widths for the second $1^+$ state, offering a means to discern between the
different interactions.

\begin{figure}[htb]
\includegraphics[angle=-90,width=1.1\figwidth]{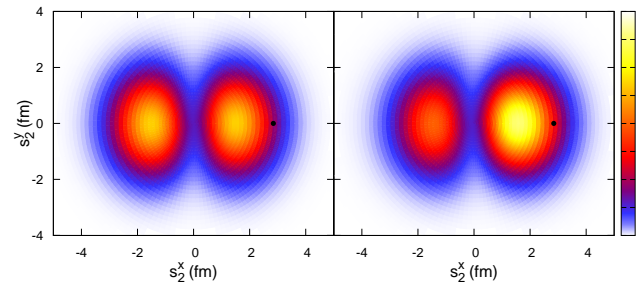}
\caption{(Color online) Projected two-nucleon density $\mathcal{P}_f(
\vec{s}_1,\vec{s}_2)$ from the two-nucleon overlap for the first $T=0$,
$1^+$ state, calculated using the NCSM and the {\sc ncsm1} Hamiltonian.
The contours show the position probability in the impact parameter plane
of nucleon 2, $\vec{s}_2$, when nucleon 1, $\vec{s}_1$, is at the position
indicated by the black point. The left panel is calculated when retaining
only the lowest $p$-shell TNA from the $N_{max}=6$ calculation. The right
panel includes the full set of NCSM TNA for $N_{max}=6$. The overlap shows
the enhanced spatial correlations arising from inclusion of single-particle
configurations of opposite parity in the two-nucleon overlap function
(see also Fig. 4 of Ref. \cite{SiT10}). The cross sections from these
two TNA sets are 6.09 and 8.52 mb, respectively. \label{fig:projden}}
\end{figure}

Further enhancement of the cross section may arise from the new couplings to
higher major shells.  Coupling to major shells of the same parity (odd $N$)
will lead to generic changes to the overall size of the two-nucleon overlap
functions and TNA. The TNA due to mixing with major shells of opposite
parity (even $N$) leads to new interference affects that can enhance two-nucleon
spatial correlations (see e.g., \cite{Pin84}). For the first $1^+$ this can
be seen in Fig. \ref{fig:projden} for calculations based on the {\sc ncsm1}
Hamiltonian. The left panel is calculated when retaining only the 0$\hbar
\omega$ $p$-shell TNA components from the $N_{max}=6$ NCSM calculation.
The right panel includes the full set of NCSM TNA for all major shells. The
enhanced spatial correlations presented to the target nucleus from the
inclusion of single-particle configurations with opposite parity in the
two-nucleon overlap function are evident. The cross sections from these
truncated and full TNA sets are 6.09 and 8.52 mb, respectively. Both
exceed those of the $p$-shell shell-model calculations, this being 3.58 mb
for the {\sc wbp} interaction TNA.

For both {\sc ncsm1} and {\sc ncsm2} the enhancement of the $np$ removal
cross sections relative to the {\sc wbp} calculations is larger than that
for the like-nucleon, $nn$, removal cross section. However, the difference
in this ratio is relatively small. Despite the larger $np$ removal cross
section obtained, some underestimation of the experimental $np$ channel
cross sections still remains. The available
data again suggests that there are remaining deficiencies in the $T=0$ parts of
the two-nucleon overlap functions. The yields to specific final states, namely
the first 3$^+$, 1$^+$ and 0$^+$ states, suggest a significant sensitivity to
the interactions used, requiring input from final-state-exclusive measurements.

\section{Summary} \label{sec:summary}
We have considered the impact of microscopic, NCSM wave function
overlaps on the theoretical cross sections for two-nucleon removal
reactions from fast \nuc{12}{C} projectiles. Data were available
for reactions on a carbon target at beam energies of 250, 1050 and
2100 MeV per nucleon.  As found in a previous analysis \cite{previous},
the $np$ removal cross sections are underestimated by the theoretical
model calculations, but do show an enhancement relative to the use
of truncated-basis $p$-shell-model calculations.  The cross sections
to both $T=0$ and $T=1$ states are enhanced, and the use of
large-basis {\sc ncsm} amplitudes does not fully resolve the relative
discrepancy between measured $np$- and $nn$-removal cross sections.

Further measurements, of final-state-exclusive cross sections and
residue momentum distributions, would allow a much more detailed
scrutiny and confrontation of the detailed reaction observables
predictions, including the identification of any indirect reaction
components arising from two-step paths to the final-states.
The calculated $np$-removal cross sections to the $T$=0, \nuc{10}{B}
final states were shown to have sensitivity to the different variants of the chiral
interactions used; for example, the ratio of the calculated
cross sections to the \nuc{10}{B} ground $3^+, T=0$ and first
$1^+, T=0$ excited states. To a lesser degree the first $0^+, T=1$
state cross section and the branching between the first $T=1, 0^+$ and
$2^+$ states was found to depend on the effective interaction.
In this case, data for the $nn$ and $pp$ removal channels would provide
useful verification.  The momentum distribution of the second $1^+, T=0$
state also shows a particular sensitivity to the interaction, providing
a further useful probe.

The overall conclusion from the present analysis is that the existing
residue-final-state-inclusive data suggest that the $T$=0, $np$-spatial
correlations present in the wave functions used are still insufficient. We
have shown that new exclusive measurements would offer a means to interrogate
these shell-model inputs, in particular for the $np$-channel, $T=0$ wave
functions, and the direct reaction mechanism predictions in considerable
detail.

\begin{acknowledgments}
This work was supported by the United Kingdom Science and Technology
Facilities Council (STFC) under Grant No. ST/J000051/1 and by the NSERC Grant No. 401945-2011.
Computing support for this work came in part from the LLNL institutional Computing Grand Challenge program.
\end{acknowledgments}

\end{document}